\documentclass{raa}

\usepackage[varg]{txfonts}

\usepackage{natbib}
\usepackage{graphicx}
\usepackage{epstopdf}
\usepackage{amssymb}
\usepackage{float}

\bibpunct{(}{)}{;}{a}{}{,}

\begin{document}

\title{The Great Pretenders Among the ULX Class}

   \volnopage{Vol.0 (200x) No.0, 000--000}      
   \setcounter{page}{1}          

\author{
Dimitris M. Christodoulou\inst{1,2},
Silas G. T. Laycock\inst{1,3},
Demosthenes Kazanas\inst{4},
Rigel Cappallo\inst{1,3},
\and
Ioannis Contopoulos\inst{5,6}
}

\institute{
Lowell Center for Space Science and Technology, University of Massachusetts Lowell, Lowell, MA, 01854, USA\\
\and
Department of Mathematical Sciences, University of Massachusetts Lowell, Lowell, MA, 01854, USA.
Email: dimitris\_christodoulou@uml.edu\\
\and
Department of Physics \& Applied Physics, University of Massachusetts Lowell, Lowell, MA, 01854, USA.
Email: silas\_laycock@uml.edu, rigelcappallo@gmail.com \\
\and
NASA Goddard Space Flight Center, Laboratory for High-Energy Astrophysics, Code 663, Greenbelt, MD 20771, USA. Email: demos.kazanas@nasa.gov \\
\and
Research Center for Astronomy and Applied Mathematics, Academy of Athens, Athens 11527, Greece.
Email: icontop@academyofathens.gr \\
\and
National Research Nuclear University, Moscow 115409, Russia. \\
}

\date{Received~~2017 month day; accepted~~2017~~month day}

\def\gsim{\mathrel{\raise.5ex\hbox{$>$}\mkern-14mu
                \lower0.6ex\hbox{$\sim$}}}

\def\lsim{\mathrel{\raise.3ex\hbox{$<$}\mkern-14mu
               \lower0.6ex\hbox{$\sim$}}}

\abstract{
The recent discoveries of pulsed X-ray emission from three ultraluminous X-ray (ULX) sources have finally enabled us to recognize a subclass within the ULX class: the great pretenders, neutron stars (NSs) that appear to emit X-ray radiation at isotropic luminosities $L_X = 7\times 10^{39}$~erg~s$^{-1}-1\times 10^{41}$~erg~s$^{-1}$ only because their emissions are strongly beamed toward our direction and our sight lines are offset by only a few degrees from their magnetic-dipole axes. The three known pretenders appear to be stronger emitters than the presumed black holes of the ULX class, such as Holmberg II \& IX X-1, IC10 X-1, and NGC300 X-1.
For these three NSs, we have adopted a single reasonable assumption, that their brightest observed outbursts unfold at the Eddington rate, and we have calculated both their propeller states and their surface magnetic-field magnitudes. We find that the results are not at all different from those recently obtained for the Magellanic Be/X-ray pulsars: the three NSs reveal modest magnetic fields of about 0.3-0.4~TG and beamed propeller-line X-ray luminosities of $\sim 10^{36-37}$~erg~s$^{-1}$, substantially below the Eddington limit. 
}

\authorrunning{Christodoulou et al.}
\titlerunning{Great Pretenders Among the ULX Class}

\maketitle


\section{Introduction}\label{intro}

Ultraluminous X-ray (ULX) sources are extragalactic compact accreting objects 
characterized by apparent super-Eddington luminosities ($L_X\sim 10^{39-41}$~erg~s$^{-1}$)  
and unusual soft X-ray spectra with blackbody emission around 
$\lesssim$ 0.3~keV and a
downturn above $\sim$3-5~keV \citep{gla09,fen11,mot14,mid15}.
The extreme luminosities observed during outbursts could be understood either
as isotropic emission below the Eddington limit 
($L_X < L_{Edd} = 1.26\times 10^{38}~M/M_\odot$~erg~s$^{-1}$) 
from intermediate-mass ($M\sim 10^{2-4} M_\odot$) black holes or 
as anisotropic emission with apparent $L_X > L_{Edd}$ from stellar-mass
black holes (BHs) and neutron stars (NSs) \citep{sor07,med13,bac14,mot14,pas14}.

The former interpretation was not supported by the results of \cite{gla09} and it is now effectively ruled out by a large number of observations: 
\begin{itemize}
\item[(a)]\cite{gil04} found that the ULX sources are just the high end of a luminosity function that cuts off at $L_X\sim 3\times 10^{40}~{\rm erg~s}^{-1}$ and in which the known high-mass X-ray binaries (HMXBs) make up the low end of a single power-law with slope $\sim$1.6.
\item[(b)]\cite{liu13} determined from optical spectroscopy that the mass of the compact object in M101 ULX-1 is no more than 30$M_\odot$. It is unlikely that this is an intermediate-mass BH.
\item[(c)]\cite{lua14} analyzed a {\it Chandra} sample of nearby ULX sources and found a change in the spectral index around $L_X\sim 2\times 10^{39}$~erg~s$^{-1}$ that may indicate a transition to the apparent super-Eddington accretion regime by 10$M_\odot$ BHs or to strongly anisotropic emission from NSs. This result was confirmed by the independent study of \cite{sut17}.
\item[(d)]\cite{bac14} determined that the ULX source X-2 in M82 harbors a pulsar with spin period $P_S = 1.3725$~s and spinup rate $\dot{P_S}= -2\times 10^{-10}$~s~s$^{-1}$.
\item[(e)]\cite{mot14} determined that the ULX source P13 in NGC7793 harbors a stellar-mass compact object with $M < 15~M_\odot$. Pulsations were next detected from this source \citep{fur16,isr16b}, so now we know that the compact object is a NS with $P_S = 0.417$~s and an average $\dot{P_S}= -3.5\times 10^{-11}$~s~s$^{-1}$.
\item[(f)]\cite{lay15} determined from the radial velocity curve of IC10 X-1 that the the compact object could be a NS, although a low-stellar-mass BH cannot be ruled out.
\item[(g)]\cite{isr16a} detected pulsations from NGC5907 ULX-1, so this is the third object of the class harboring a NS with $P_S = 1.137$~s and an average $\dot{P_S}= -8.1\times 10^{-10}$~s~s$^{-1}$. This is also the most luminous and the most distant NS pretender ever detected, given its apparent luminosity of $L_X = 10^{41}$~erg~s$^{-1}$ and a distance to the source of $D=$17.1~Mpc.
\end{itemize}

The magnetic field of NGC7793 P13 has been recently estimated to be $B\approx 1.5$~TG \citep{fur16}, and in the case
of M82 X-2, \cite{bac14} used their measurement of the accretion torque
to obtain a modest value of the magnetic field
$B\approx 1$~TG. Somewhat higher values were obtained by \cite{isr16a} for the magnetic field of NGC5907 ULX-1 in which these authors also analyzed the possibility of multipolar fields on the surface of the NS \citep[see also][for a similar analysis in the case of M82 X-2]{che17}. As we shall see below, such assumptions may not be necessary as the faint X-ray states and the magnetic fields of these NSs are determined to be quite modest and in strong agreement with the estimates of \cite{kin16} for M82 X-2 and with the estimates of \cite{chr16} for several Magellanic HMXBs.
 
More evidence keeps piling up that the magnetic fields of the ULX objects are not exotic. For instance, \cite{bri16},
based on {\it NuSTAR} observations of M82 X-2, found
a spectral cutoff at $14_{-3}^{+5}$~keV which implies a 1.2~TG surface magnetic field. On the other hand, \cite{tsy16} and \cite{dal16} have interpreted the faintest states of M82 X-2 as low-level
emission that occurs due to leakage and continued accretion of matter when the system moves into
the \cite{cor96} gap. As we shall see below, this interpretation that leads to higher magnetic-field values is not supported by the theory or the observations of the other two pretenders.

In \S\ref{calc}, we calculate the physical properties of the faintest accreting states of the three NS pretenders (their so-called ``propeller'' states) that lie at anisotropic X-ray luminosities of $\sim 10^{36-37}$~erg~s$^{-1}$. We adopt a single reasonable assumption, that the brightest radiation seen from these NSs is beamed emission that proceeds at the Eddington rate\footnote{This assumption is supported by the recent simulations of anisotropic outflows from NS accretion columns by \cite{kaw16}.\label{ft1}} (despite the observed spread of more than an order of magnitude in the highest X-ray luminosities of these sources). In that respect, the outbursts of these sources are not at all dissimilar from the type-II outbursts of Magellanic Be/X-ray pulsars that have been observed to rise up to the Eddington limit \citep[][their Figures~2 and 4]{coe10,chr16}. Then we show that at their faintest accreting states, these objects appear to be very similar and typical NSs; emitting anisotropically modest amounts of radiation; and supporting modest surface magnetic fields; properties that are quite similar to those also found for the propeller emission states of the Magellanic Be/X-ray pulsars.
In \S\ref{conclusions}, we discuss and summarize our results.

%
\begin{table}
\caption{NS ULX Measurements}
\label{t1}
\centering
\begin{tabular}{lccccl}
\hline\hline
Source & $L_{max}$ & $L_{min}$  & $P_S$ & $\dot{P_S}$ & $D$ \\
Name & (erg/s) & (erg/s)  & (s) & (s s$^{-1}$) & (Mpc) \\
\hline
NGC7793 P13  & $7\times 10^{39}$ & $5\times 10^{37}$ &  0.417 & $-3.5\times 10^{-11}$ & 3.6 {\rm or} 3.9\\
NGC5907 ULX-1  & $1\times 10^{41}$ & $3\times 10^{38}$ &  1.137 & $-8.1\times 10^{-10}$ & 17.1\\
M82 X-2  & $2\times 10^{40}$ & $1\times 10^{38}$ &  1.3725 & $-2.0\times 10^{-10}$ & 3.6 \\

\hline
\end{tabular}
\\
References.---NGC7793 P13: \cite{mot14}, \cite{fur16}, \cite{isr16b}. NGC5907 ULX-1: \cite{isr16a}. M82 X-2: \cite{bac14}, \cite{bri16}. 

\end{table}

%
\begin{table}
\caption{NS ULX Estimates}
\label{t2}
\centering
\begin{tabular}{lcccl}
\hline\hline
                          & NGC7793 & NGC5907 & M82 &  \\ 
Source Property  & P13         & ULX-1      &  X-2  & Input Parameters \\ 
\hline
$L_{max}$(10$^{40}$~erg/s)       &  0.7   & 10   & 2      & Values from Table~\ref{t1}\\
$L_{Edd}$(10$^{38}$~erg/s)        &  1.8   & 1.8  & 1.8   & For $M=1.4~M_\odot$\\
Beaming Factor $b$                        &  38.9 & 556 & 111  & $=L_{max}$/$L_{Edd}$, eq.~(\ref{b})\\
Half-Opening Angle $\theta_{\frac{1}{2}}$($^o$) &  18    &  5.0 &  11   & $=114.6/\sqrt{b}$, eq.~(\ref{theta}) \\ 
\hline
$L_{iso}$(10$^{39}$~erg/s)    & 0.50  & 1.1  & 0.18 & $P_S$ and $\dot{P_S}$ in eq.~(\ref{liso}) with $\eta = 0.5$ \\
$L_{prop}$(10$^{36}$~erg/s) & 13     & 2.0  & 1.6   & $=L_{iso}/b$, eq.~(\ref{Laniso}) \\
$B$ (TG)                                    & 0.29  & 0.37& 0.41 & $P_S$ and $L_{prop}$ in eqs.~(\ref{stella}) and (\ref{mu}) \\
\hline
$L_{min}/b$ (10$^{36}$~erg/s) & 1.3     & 0.54  & 0.90   & $L_{min}$ from Table~\ref{t1} and $b$ \\
\hline
\end{tabular}
\\
Error Bars.---Not accounting for errors in the distances to the sources, the errors determined from the observations in Table~\ref{t1} are as follows (left to right): $\Delta(\ln P_S) = 4.8\times 10^{-6}, 3.5\times 10^{-6}, 8.7\times 10^{-8}$; $\Delta(\ln\dot{P_S}) = 8.6\times 10^{-4}, 1.2\times 10^{-2}, 6.9\times 10^{-3}$; and $\Delta(\ln L_{max}) = 0.015, 0.20, 0.12$. The calculated errors are as follows (left to right): $\Delta(\ln L_{iso}) = 8.6\times 10^{-4}, 1.2\times 10^{-2}, 6.9\times 10^{-3}$; and $\Delta(\ln b)=\Delta(\ln[L_{min}/b]) = \Delta(\ln L_{prop}) =0.015, 0.20, 0.12$.

\end{table}

\section{The Propeller States of the Three Pretenders}\label{calc}

In Table~\ref{t1}, we summarize the recent measurements of the properties of the three NS pretenders. Two trends appear to be common for the group: 
\begin{itemize}
\item[(1)]The NSs are spinning up over times that span 18 days (M82 X-2) to 4-12 years \citep[12 years in the case of NGC5907 ULX-1;][]{isr16a}. This indicates that accretion of matter with high specific angular momentum spins up these pulsars while, at the same time, changes in the magnetospheres (such as the opening of magnetic-field lines and the electric currents flowing along open lines) are not able to generate strong enough retarding torques\footnote{Using eq.~(18) in \cite{con14}, the measured values of $P_S$, and the below-determined values of $B$, we find that the spin-down rates for the pretenders are $\dot{P}_{SD} = 0.82\pi^2 B^2 R^6/(c^3 I P_S)\sim 10^{-17}$~s~s$^{-1}$, values that are clearly negligible as compared to the measured spin-up rates listed in Table~\ref{t1}. Here $c$ is the speed of light and $R$ and $I$ are the canonical values of the NS radius and moment of inertia, respectively. For pulsars such as the pretenders with $P_S\sim 1$~s, this estimate can be increased by a factor of $(R_{lc}/R_{co})^2\sim 10^3$ depending on the locations of the corotation radius $R_{co}$ and the light-cylinder radius $R_{lc}$ \citep{con05}, but still the resulting spin-down rates are orders of magnitude smaller than the observed spin-up rates. It appears then that in pulsars with $P_S\sim 1$~s, the inward push of the accretion disks cannot disrupt the magnetospheres enough to affect strong electromagnetic spin-down. 
}
to reverse this secular trend \citep{har99,con99,con14}. 
\item[(2)]The lowest observed X-ray luminosities $L_{min}$ are deceptively close to the Eddington limit for the canonical $1.4M_\odot$ pulsar. But we cannot trust this trend to be more than a coincidence because we do not know whether these $L_{min}$ values represent faint accreting states or faint magnetospheric emission \citep{cam95,cam97}, since pulsations have not been detected in these observations. The nature of these states can be revealed after the nature of the brightest states $L_{max}$ is deciphered (see below).
\end{itemize}

We carry out a sequence of calculations that we also summarize in the rows of Table~\ref{t2}. We begin with the extreme values of the observed X-ray luminosities $L_{max}$ (Table~\ref{t1}) which we introduce in the first row of Table~\ref{t2}. As many other researchers have stated in the past, we believe that these luminosities cannot indicate isotropic emissions from these pulsars---instead, the radiation must be strongly beamed toward our direction. But we do realize that the sources displaying these $L_{max}$ values are undergoing powerful collimated outbursts, in which case it is reasonable to assume that the emitted radiation is limited by the Eddington rate in all cases (see also footnote \ref{ft1}). We also adopt canonical pulsar parameters (mass $M=1.4 M_\odot$ and radius $R=10$~km) throughout the derivations. Then $L_{Edd} = 1.8\times 10^{38}$~erg~s$^{-1}$, 
and we can calculate the beaming factors of the emitted X-rays which we define here as
\begin{equation}
b \equiv \frac{L_{max}}{L_{Edd}} > 1 \, .
\label{b}
\end{equation}
The beaming factors for critical accretion give us an idea about the half-opening angles $\theta_{\frac{1}{2}}$ of the collimated emissions. The solid angles are $4\pi /b$ steradians and then the half-opening angles of the emission cones $\theta_{\frac{1}{2}}$ are given by $\pi(\theta_{\frac{1}{2}})^2 = 4\pi / b$, viz.
\begin{equation}
\theta_{\frac{1}{2}} = \left(\frac{360}{\pi}\right)b^{-1/2}~{\rm deg.}
\label{theta}
\end{equation}
We find that $\theta_{\frac{1}{2}}$ varies between 5 and 18 degrees, where the most stringent value corresponds to the most luminous outburst (NGC5907 ULX-1 in Table~\ref{t2}). This value is in agreement
with the range of $\theta_{\frac{1}{2}}=5.0$-6.7 degrees found for this source by \cite{dau17} who produced a model with a slightly precessing inflow and outflow and strong beaming. The accretion in this model is however supercritical and the maximum anisotropic luminosity emitted in the funnel is found to be in the range of (5-8)$L_{Edd}$. Nevertheless, the model of \cite{dau17} is important because it shows that a slight precession of the outflow can reproduce the nearly sinusoidal pulse profiles observed in the NS pretenders. The shape of the pulse profiles was used in the past to raise objections against strong beaming \citep[][among others]{bac14,bri16,fur16}, but now it does not appear to be a problem. 

\begin{figure}
\includegraphics[scale=1]{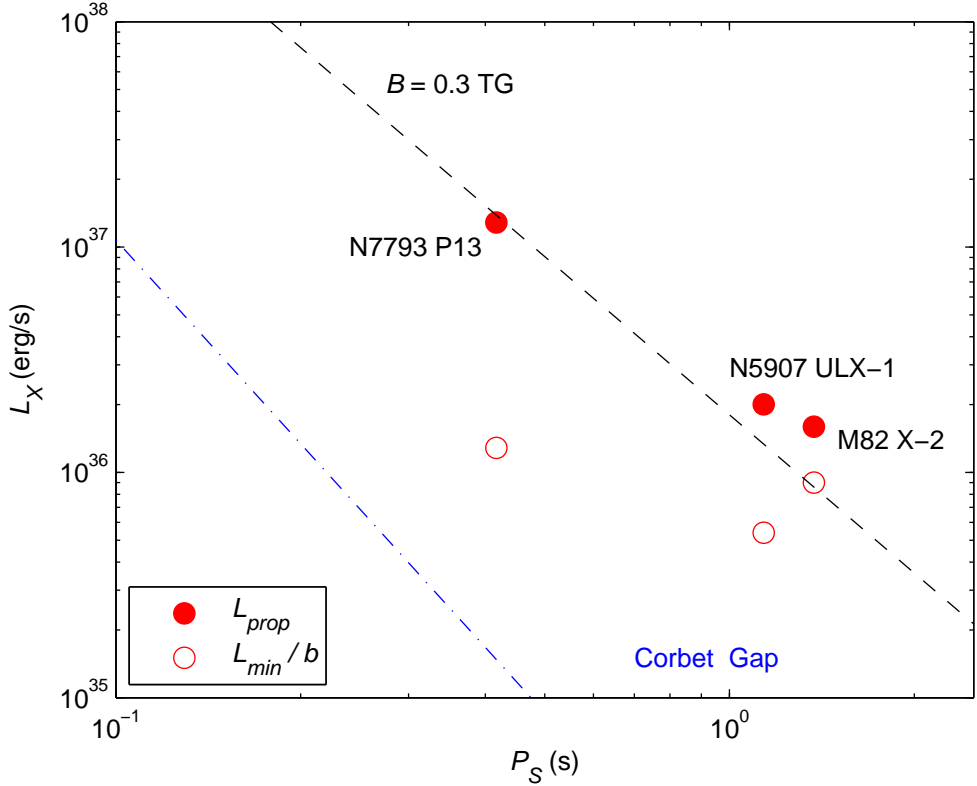}
\caption{The anisotropic propeller states (i.e., the faintest accreting states; solid circles) and the faintest anisotropic states observed (open circles) in $P_S-L_X$ space for the three NS pretenders. The data points come from Tables~\ref{t1} and~\ref{t2}. Errors are given in Table~\ref{t2}; error bars are very small for $P_S$ and about the size of the circles for $L_X$ (smaller for NGC7793 P13). The dashed line is the lowest propeller line with $B=0.3$~TG found for the Magellanic Be/X-ray pulsars \citep{chr16}. Comparison indicates that the surface magnetic fields of these NSs are modest ($B\approx 0.3$-0.4~TG). The dashed-dotted line with a slope of $-3$ specifies the lower boundary of the \protect\cite{cor96} gap for the $B=0.3$~TG propeller line.}
\label{fig1}
\end{figure}

Next we use the measured values of $P_S$ and $\dot{P_S}$ (Table~\ref{t1}) to find the isotropic X-ray luminosities $L_{iso}$ of the NSs in their faintest accreting states \citep[Appendix B in][]{gal08,fra02}, viz.
\begin{equation}
L_{iso} = \frac{1}{2}\eta\left(2\pi I |\dot{P_S}|\right)\left( \frac{2\pi}{P_S^7}\frac{G M}{R^3} \right)^{1/3} \, ,
\label{liso}
\end{equation}
where $I$ is the canonical moment of inertia, $G$ is the gravitational constant, and we introduced an additional  factor of $\eta /2$ for the efficiency of converting accretion power to X-rays ($\eta$ taken here to be 0.5). The leading factor of 1/2 applies specifically to collimated outflows from NSs \citep{kor06} and it implies an effective NS conversion efficiency of $\eta/2=0.25$. The introduction of $\eta/2$ also implies that we assume that minimum accretion takes place at a reduced torque than the observed maximum value. In the case of M82 X-2, this is certainly true because the observed value of $\dot{P_S}$ was obtained only at outburst and for a very limited time span of 18 days \citep[Fig.~2a in][]{bac14}.

Next the isotropic ``propeller'' luminosities $L_{iso}$ are scaled by $b$ and the true anisotropic X-ray luminosities 
\begin{equation}
L_{prop} = \frac{L_{iso}}{b} \, ,
\label{Laniso}
\end{equation}
are obtained for collimated emission in the propeller states (see Table~\ref{t2}). The faintest observed isotropic luminosities $L_{min}$ shown in Table~\ref{t1} are also scaled by $b$ and they are introduced at the bottom of Table~\ref{t2} for comparison purposes. 

Finally, a minimum value for the surface magnetic field $B$ is estimated from the standard equation of \cite{ste86} for the propeller line, viz.
\begin{equation}
L_{prop} = 2\times 10^{37} \left(\frac{\mu}{10^{30}~{\rm G~cm^3}}\right)^2 
\left(\frac{P_S}{1~{\rm s}}\right)^{-7/3} ~{\rm erg~s^{-1}}\, ,
\label{stella}
\end{equation}
where canonical pulsar parameters have been used and the magnetic moment is defined by
\begin{equation}
\mu\equiv B R^3 \, . 
\label{mu}
\end{equation}
This equation does not depend on $\dot{P_S}$, thus it does not rely on torque balance at the inner edge of the accretion disk and it does not use the conservation of angular momentum \citep{fra02} in the determination of the magnetic field.

The results for $L_{prop}$ summarized in Table~\ref{t2} are plotted in the $P_S-L_{X}$ diagram in Figure~\ref{fig1} as filled circles. We see now that these three pulsars are totally nonexotic, sporting propeller luminosities in the range of  $L_{prop}\sim 10^{36-37}$~erg~s$^{-1}$ and surface magnetic fields in the range of $B\approx 0.3$-0.4~TG. In these respects, the pretenders are not at all dissimilar from the short-period HMXBs found in the Magellanic Clouds and studied under the assumption of isotropic emission \citep{chr16}. In fact, several of the Magellanic sources also appear to rise up during powerful type-II outbursts to just about the Eddington limit, an observation that is responsible for our adoption of this assumption in the present work.

The $B$-values found for the three pretenders are in agreement with the results obtained by \cite{kin16} for M82 X-2 and other accreting NSs that exhibit beamed emission at the apparent level of $L_{max}\sim 10^{40}$~erg~s$^{-1}$. The importance of our results is that they predict similar modest values and properties for NS ULX sources that appear to radiate at a much higher level of power (see NGC5907 ULX-1 in Table~\ref{t2}).

The faintest anisotropic states $L_{min}/b$ of the three pretenders shown in Table~\ref{t2} are also plotted in Figure~\ref{fig1} as open circles. For the measured spin periods, these states fall within the \cite{cor96} gap which suggests that we have observed weak magnetospheric emission in the absence of accretion \citep{cam95,cam97}. The case for NGC7793 P13 is especially strong: the faintest observation  \citep{mot14} lies one order of magnitude below the propeller value of $L_{prop} = 1.3\times 10^{37}$~erg~s$^{-1}$. On the other hand, the faintest observation of M82 X-2 \citep{bri16} lies only a factor of $\sim$2 below the propeller value $L_{prop} = 1.6\times 10^{36}$~erg~s$^{-1}$. But for this pulsar there exists an independent confirmation of its propeller value: From 15 years of {\it Chandra} observations, \cite{tsy16} found that M82 X-2 has repeatedly switched between its high isotropic state $L_{max}$ and a low isotropic state with $L_{iso}= 1.7\times 10^{38}$~erg~s$^{-1}$ \citep[see also][who obtained the same value as an upper limit]{dal16}. Using $b=111$ (Table~\ref{t2}), this faint state corresponds to $L_{prop} = 1.5\times 10^{36}$~erg~s$^{-1}$ in strong agreement with our determination of the propeller value derived from beaming of the $L_{max}$ value of this source. It seems then that M82 X-2 has bounced for many years between its propeller state and its ultraluminous state. This agreement between results consolidates the physical properties of M82 X-2 listed in Table~\ref{t2}.

\section{Summary and Discussion}\label{conclusions}

We have used standard accretion theory and beaming of the X-ray emission \citep{ste86,fra02,gal08,kin16} in order to estimate the typical values of the minimum anisotropic luminosities and the surface magnetic fields of the three recently discovered ULX pulsars \citep[Table~\ref{t1};][]{bac14,fur16,isr16a,isr16b}. For the measured values of $L_{max}$, $P_S$, and $\dot{P_S}$, our results show that the physical and geometric parameters are all modest (Table~\ref{t2}) and not at all dissimilar (Fig.~\ref{fig1}) from those found for the entire sample of the short-period ($P_S < 100$~s) Magellanic HMXBs \citep{chr16}. The three modest NSs only pretend to emit at enormous super-Eddington rates because their emissions are collimated and our sight lines are close to their magnetic-dipole axes (\S~\ref{calc}). 

Since the X-rays originate from a limited region around the magnetic poles, we can estimate the sizes $r$ and temperatures $T$ of these hot spots during the faintest accretion states in which the corotation radius is comparable to the magnetospheric radius. The radii of the hot spots are approximately given by $r = R\sqrt{R/R_{co}}$, where $R_{co}$ is the corotation radius; and then the temperatures are $T = [L_{Edd}/(\sigma\pi r^2)]^{1/4}$, where $\sigma$ is the Stefan-Boltzmann constant. Using the values listed in Table~\ref{t1}, we find that typically $r = 1.0-0.7$~km and $T= (1.0-1.2)\times 10^8$~K corresponding to photon energies of $kT= 8.5-10.4$~keV at the bases of the outflows. Such temperatures may be sufficient for the production of particle pairs that can cross magnetic-field lines and escape along the magnetic axes in mildly relativistic outflows that help increase the effective opening angles of the emission cones listed in Table~\ref{t2}.

Furthermore, if the emissions from the pretenders are collimated, some of the energy must emerge at much longer wavelengths. This appears to be the case for M82 X-2 according to the radio observations of \cite{kro85}, \cite{mcd02}, and \cite{fen08};
and the infrared observations of \cite{kon07} and \cite{gan11}. 
The radio maps show a core-dominated source which is expected if the pulsar is a 
modestly aligned rotator and a collimated jet is coming out in our direction. 
\cite{kon07} also produced {\it Chandra} X-ray spectra that are
hard (photon indices $1.3-1.7$ from an absorbed power-law model) and show no soft excess.
This, combined with the strong X-ray variability on timescales of $\sim$2 months and the 
reccuring type-II outbursts, indicates that M82 X-2 is not at all dissimilar from Galactic and
Magellanic X-ray binaries harboring NSs \citep[][]{yang17}.
The recent X-ray observations reported by \cite{bri16} and \cite{tsy16}
for M82 X-2 have effectively confirmed this picture.

In the Magellanic Clouds, two HMXBs have been observed each during two major type-II outbursts\footnote{A type-II outburst of SMC X-3 was recently reported by \cite{wen17} and \cite{tsy17} and it was monitored by {\it Swift/XRT} and {\it NuSTAR}. For this HMXB, the observations indicate that $L_{max}=2.5\times 10^{39}~{\rm erg~s}^{-1}$ and $L_{min}=3.0\times 10^{34}~{\rm erg~s}^{-1}$ \citep[$D=62$~kpc;][]{tsy17}. Assuming that this event was due to critical accretion at the Eddington rate, we find that we need modest beaming ($b=14$ and $\theta_{\frac{1}{2}} = 31^o$), the minimum beamed luminosity $L_{min}/b$ falls in the middle of the Corbet gap, and the dipolar magnetic field on the surface of the NS is $B=1.1$~TG. 
} with apparent $L_{max} > L_{Edd}$: LXP8.04 \citep[][]{edg04,vas14,ten14} and SMC X-2 \citep[][]{lay05,ken15}. As summarized by \cite{chr16}, the brightest bursts reached luminosities of $L_{max} = 8\times 10^{38}~{\rm erg~s}^{-1}$ (LXP8.04, $D=50$~kpc) and $L_{max} = 4\times 10^{38}~{\rm erg~s}^{-1}$ (SMC X-2, $D=60$~kpc) that reveal small beaming factors (eq.~[\ref{b}]) of about 2.2 and 4.4, respectively. These very small degrees of beaming (half-opening angles of $\theta_{\frac{1}{2}} = 77^o$ and 54$^o$, respectively; eq.~[\ref{theta}]) and our orientation within such wide emission cones account for the differences in the light curves and the spectral features between these most extreme HMXBs and the strongly beaming NS pretenders. The ULX spectral features are discussed below.

The main properties of the so-called ultraluminous state \citep{gla09,fen11,mot14,mid15,kob17} are:
\begin{itemize}
\item[1.]A soft blackbody (BB) excess at $\lesssim$ 0.3 keV. It is believed that this could be emission from the accretion disk.
\item[2.]A downturn of the spectrum at $\sim$3-5 keV. This is consistent with our calculation of photon energies of $kT= 8.5-10.4$~keV at the bases of the outflows. Operating at $L_{max}/b=L_{Edd}$, the sources would not be able to produce more energetic photons in substantial numbers. A large number of photons is also capable of escaping from the sides of the accretion column \citep{kaw16,bas76} in a fan-like configuration and most of them do not reach the observer if our sight lines are oriented close to the magnetic-dipole axes, as in the case of the pretenders.
\item[3.]Featureless spectra with no emission/absorption lines. This implies that the X-rays are not reprocessed in the surrounding medium, that is they find holes in a clumpy medium to shine through \citep[as was found by][in {\it Suzaku} observations of Holmberg IX X-1]{kob17}.
\item[4.]A power law at the hard part of the spectrum. Comptonization of the primary (unprocessed) photons from the source may be responsible for this component.
\item[5.]The soft part is often fitted well by two-color disk BB models. Multi-color BB models could be revealing the spectrum of a cooler wind outflow emanating in the emission funnel \citep[see also][]{wal16}.
\item[6.]No cyclotron resonance features. This can be explained by the weak magnetic fields of the pretenders. For the determined values of $B = 0.3$-0.4~TG, such lines may only emerge at 3.5-4.6~keV which is the area of the spectrum downturn.
\end{itemize}

\cite{urq16} observed eclipses in two ULX sources in M51. For these sources, $L_{max}\approx 2\times 10^{39}$~erg~s$^{-1}$, substantially lower than the $L_{max}$ values of the pretenders and slightly below the empirical critical value of $L_{crit}\sim 3\times 10^{39}$~erg~s$^{-1}$ \citep{lua14,sut17} that apparently subdivides the central engines of the ULX sources into stellar-mass BHs and NS pretenders. Since our line of sight is very much inclined to the magnetic axes of both of these eclipsing binaries, the radiation cannot be beamed at all. We conclude that both of these sources contain stellar-mass BHs with masses $M_{BH}\gtrsim 16M_\odot$ radiating isotropically at about the Eddington rate ($L_{max}\lesssim L_{Edd}$). Objects such as these reinforce our belief that the NSs of the ULX class only pretend to radiate more power than the BHs of the class. In this respect, weaker sources with $L_{max}\sim 2\times 10^{38}$~erg~s$^{-1}$, such as IC10 X-1 \citep{lay15} and NGC300 X-1 \citep{bin11}, that bridge the gap between HMXBs and ULX sources may not contain BHs (their masses would have to be no more than $2M_\odot$), so they could also contain NSs; otherwise most of the radiation is beamed away from our direction (IC10 X-1 is an eclipsing binary); or accretion is markedly suppressed in these objects, perhaps for some of the reasons recently put forth by \cite{tut16}.

\begin{acknowledgements}
We thank an anonymous referee for suggestions that led to improvements and clarifications in the paper. DMC, SGTL, and RC were supported by NASA grant NNX14-AF77G. DK was supported by a NASA ADAP grant.
\end{acknowledgements}

\label{lastpage}

\end{document}